\documentclass[twoside]{ilcws07}
\usepackage[latin1]{inputenc}
\usepackage[dvips]{graphicx,epsfig,color}
\usepackage{wrapfig,rotating}
\usepackage{amssymb,amsmath,array}
\usepackage[colorlinks,dvips,ps2pdf]{hyperref}

\begin{document}
\title{
The LCFIVertex Package: vertex detector-based Reconstruction at the ILC} 
\author{S. Hillert on behalf of the LCFI collaboration
\vspace{.3cm}\\
University of Oxford - Department of Physics \\
Denys Wilkinson Building, Keble Road, Oxford OX1 3RH - UK
}

\maketitle

\begin{abstract}
The contribution gives an overview of the LCFIVertex package, providing 
software tools for high-level event reconstruction at the International 
Linear Collider using vertex-detector information. The package was 
validated using a fast Monte Carlo simulation. Performance obtained with 
a more realistic GEANT4-based detector simulation and realistic tracking 
code is presented. The influence of hadronic interactions on flavour 
tagging is discussed.
\end{abstract}

\section{Introduction}

At the International Linear Collider (ILC), the vertex detector is expected 
to provide high-precision measurements with a point resolution of   
$\sim3\,\mu\mathrm{m}$ or below over an unprecedented acceptance region of 
$\left|\cos\theta\right| < 0.98$, where $\theta$ is the track polar angle. 
The ambitious detector design will result in excellent vertexing and flavour 
tagging performance. This contribution \cite{url} describes scope, 
validation and resulting performance of the LCFIVertex package, a software 
package for vertex detector-based event reconstruction at the ILC.

The LCFIVertex package provides tools for vertexing, flavour tagging and 
the determination of the heavy quark charge sign of the leading hadron in 
heavy flavour jets. For vertexing, the ZVTOP topological vertex finding 
algorithm developed by D. Jackson for SLD is implemented 
\cite{Jackson:1996sy}. For the first time within an ILC software environment, 
the more specific ZVKIN branch of the algorithm is provided in addition to 
the widely used ZVRES branch. 

The ZVRES approach uses a vertex function calculated from ``probability 
tubes'' representing the tracks in a jet. Maxima of the vertex function in 
three-dimensional space are sought and the $\chi^{2}$ of the vertex fit 
is minimised by an iterative procedure. In contrast, ZVKIN initially 
determines the best approximation to the direction of flight of the 
$B$-hadron in candidate $b$-jets and uses the additional kinematic 
information provided by this ``ghost track'' to find 1-prong decay vertices 
and short-lived $B$-hadron decays not resolved from the interaction point.

Flavour tagging is accomplished using a neural net approach. The software 
includes a full neural network package written by D. Bailey. By default, 
the flavour tag is obtained from the algorithm developed by R. Hawkings 
\cite{Xella Hansen:2000dr}, however it should be noted that the 
implementation is highly flexible, allowing the user to change the input 
variables for the network, its architecture and training method.

Determination of the quark charge is initially limited to cases in which 
the leading hadron is charged using an approach developed at SLD and 
modified as described at a previous LCWS workshop \cite{Hillert:2005rp}. 
With ZVKIN, the basis is provided for extending the functionality to cover 
also neutral hadrons, using the SLD charge dipole technique 
\cite{Thom Wright}.

The C++-based package uses the LCIO data format \cite{Gaede:2003ip} for 
input and output and is interfaced to the analysis and reconstruction 
framework MarlinReco \cite{Wendt:2007iw}. An interface to the JAS3-based 
US software framework org.lcsim \cite{org.lcsim} is planned to be written 
in the near future. The code is available at the ILC software portal 
\cite{ILC software portal} and the Zeuthen CVS repository 
\cite{Zeuthen CVS}. To facilitate comparisons of results from ILC physics 
studies performed by different groups, in addition a new CVS repository 
called ``tagnet'' has been set up in Zeuthen for providing trained neural 
networks in the format used by this package. Users training their own 
neural networks, e.g. for specific physics processes, are encouraged to 
make them available to the community in this way.

\section{Code validation}
%
\begin{figure*}[h]
\centering
\mbox{
\includegraphics[width=0.80\columnwidth]{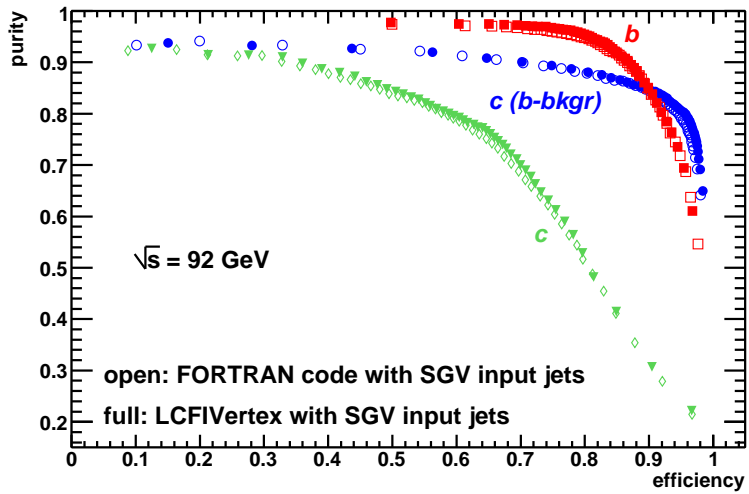}}
\caption{Comparison of tagging performance achieved with the LCFIVertex 
package and the previous FORTRAN code, using identical input events at the 
$Z$-resonance. Tagging purity is shown as function of efficiency for $b$-jets 
and $c$-jets. Performance for $c$-jets assuming only $b$-background is also 
shown. All tagging results are in excellent agreement.}
\label{Fig:Validation}
\end{figure*}
Prior to development of the LCFIVertex package, part of its functionality
was available in the form of FORTRAN routines which had been used in
conjunction with the BRAHMS Monte Carlo (MC) simulation and reconstruction
\cite{Behnke:2001se} used for the TESLA-TDR \cite{Behnke:2001qq}. 
The fast MC program Simulation a Grande Vitesse, SGV, developed by 
M. Berggren \cite{SGV} and interfaced to this FORTRAN flavour tag 
\cite{Adler:2006dw}, allowed detailed cross checks to be performed during 
the development phase. Results from tests of various separate parts of the
package were presented at the 2006 ECFA ILC workshop \cite{Jeffery}.

As final step of the code validation, the resulting flavour tagging
performance achieved with the FORTRAN/SGV code and with our package was
compared, using identical $e^{+}e^{-} \rightarrow q\bar{q}$ events
$(q = u,\,d,\,s,\,c,\,b)$ at the $Z$-resonance and at a centre of mass
energy of $\sqrt{s} = 500\,\mathrm{GeV}$, generated with the PYTHIA MC
program \cite{Sjostrand:2006za}. The SGV MC simulation was extended to 
write out the relevant information in LCIO format as input to our code 
using the FORTRAN interface provided within LCIO.

Figure~\ref{Fig:Validation} shows the resulting tagging performance from the 
LCFIVertex package compared to that of the FORTRAN code in terms of tagging 
purity vs efficiency, for events at the $Z$-resonance. Excellent agreement 
is seen for the $b$- and the $c$-tag as well as for the $c$-tag when 
considering $b$-background only, as is relevant for some physics processes. 
Agreement at the higher energy is equally good.
%
%
\begin{figure*}[h!]
\centering
\begin{tabular}{cc}
\mbox{\includegraphics[width=0.45\columnwidth]{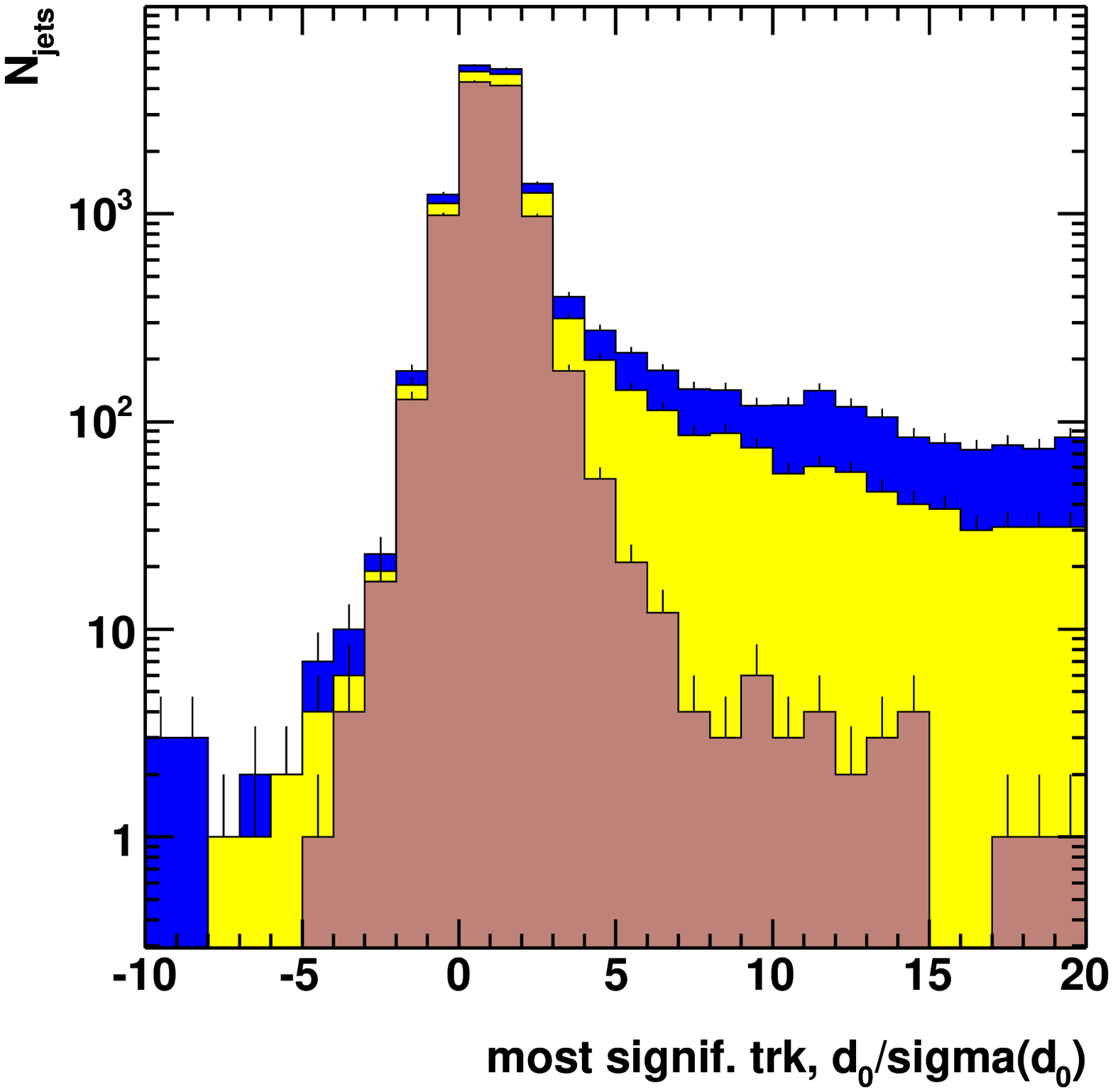}} &
\mbox{\includegraphics[width=0.45\columnwidth]{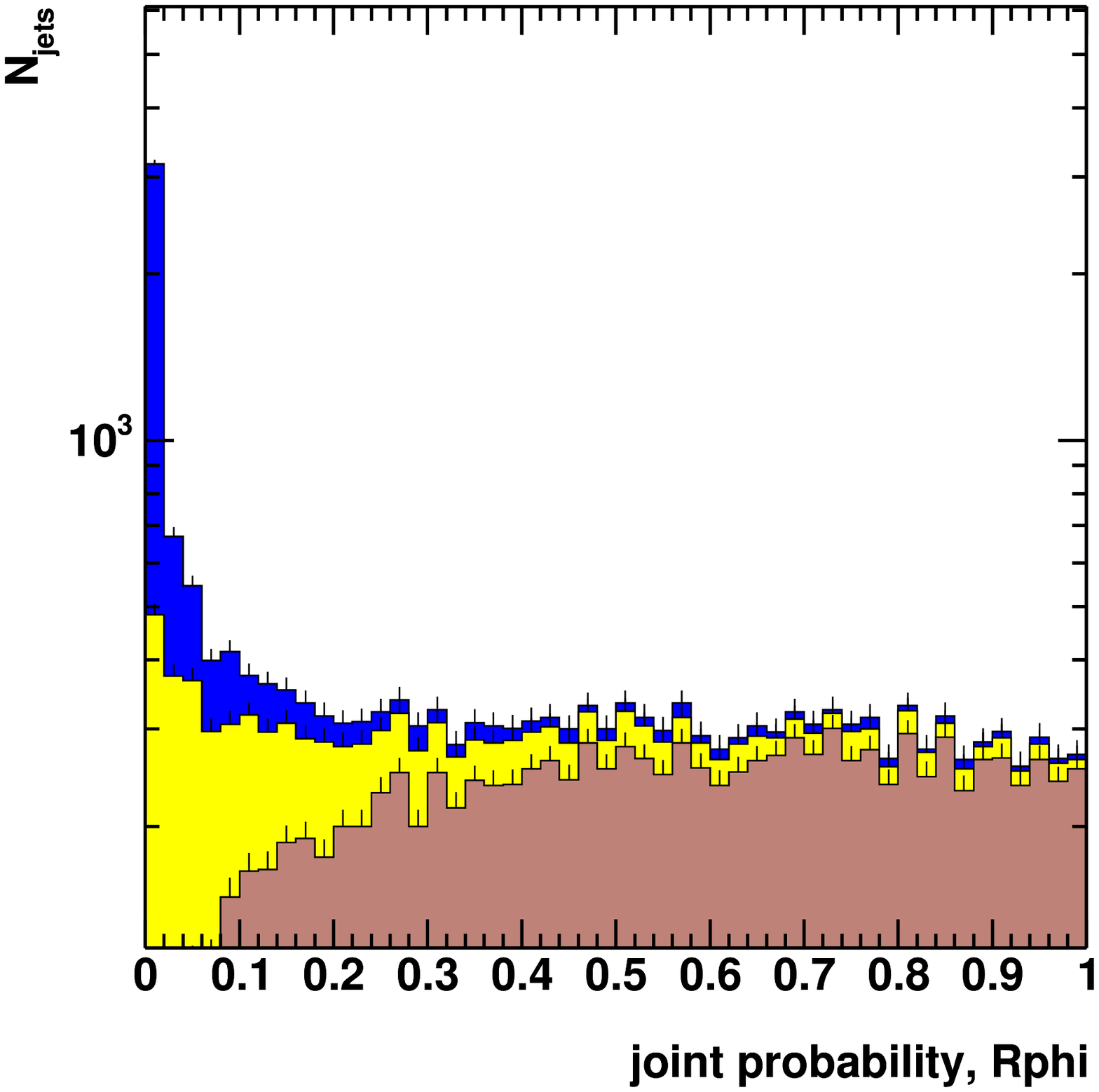}}\\
(a) & (b)\\
\mbox{\includegraphics[width=0.45\columnwidth]{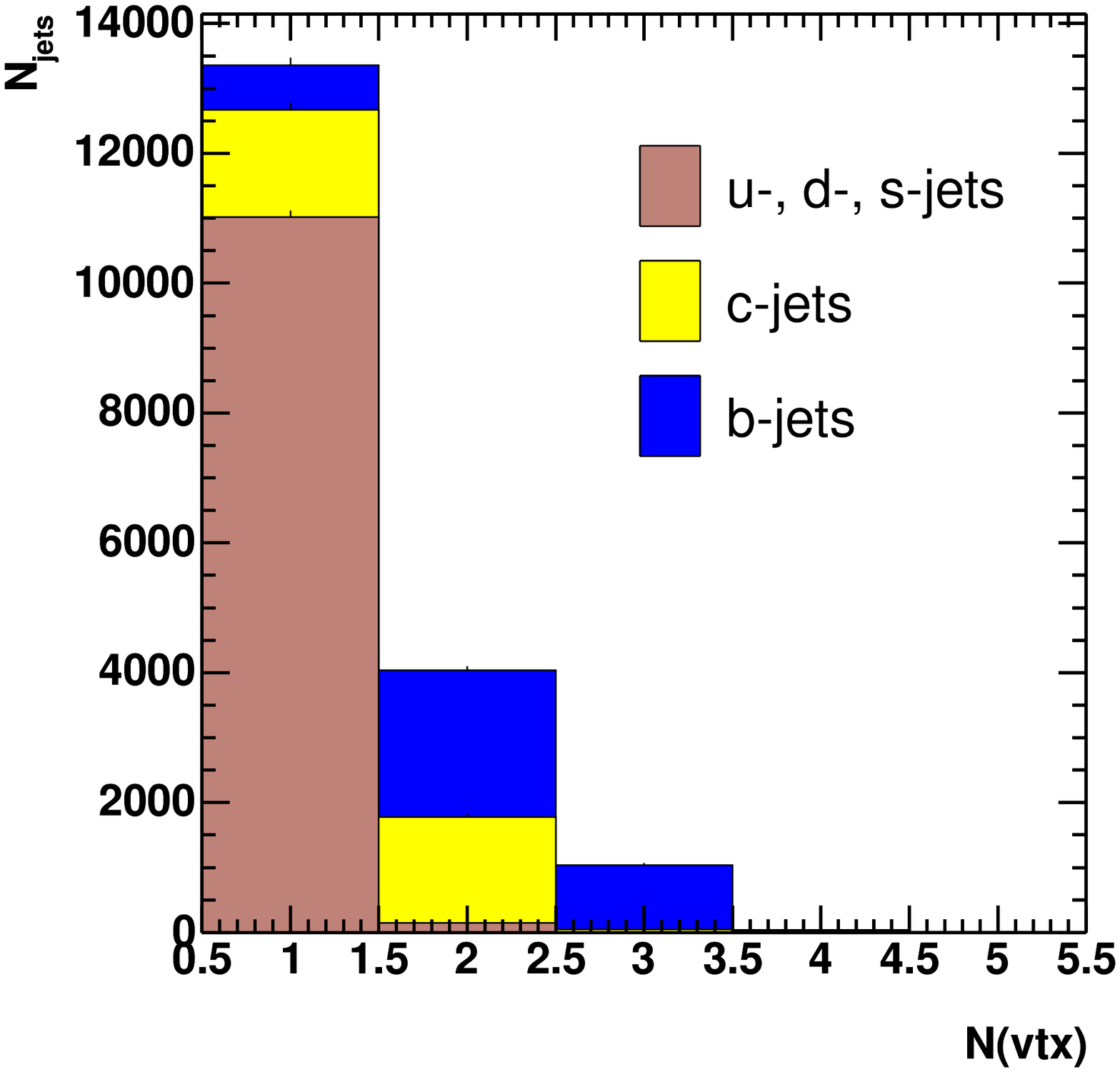}} &
\mbox{\includegraphics[width=0.45\columnwidth]{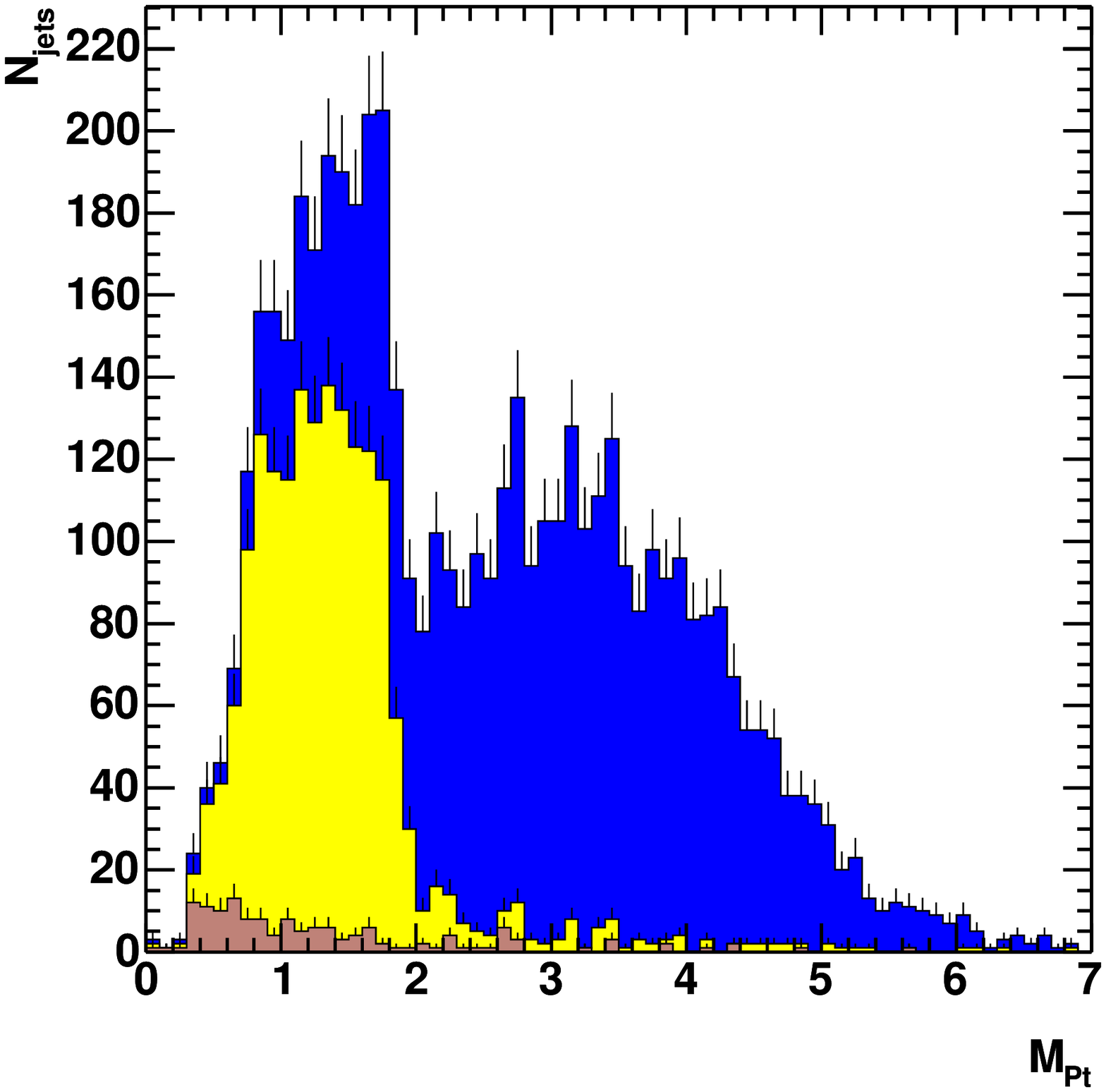}}\\
(c) & (d)\\
\end{tabular}
\caption{Some of the variables used to distinguish the different jet flavours. 
Shown are (a) the impact parameter signficicance of the most significant track 
in the jet, (b) the ``joint probability'' for all tracks in the jet to 
originate at the event vertex, (c) the vertex multiplicity in the jet and (d) 
the $P_{t}$-corrected vertex mass.}
\label{Fig:FlavourTagInputs}
\end{figure*}
\section{Performance obtained with full MC}
Following the successful code validation, performance of the LCFIVertex
package with more realistic input was studied. For this purpose, the same
PYTHIA events used for the validation were passed through the GEANT4-based
full MC simulation MOKKA, version 06-03 \cite{Mora de Freitas:2004sq}, 
assuming the detector model LDC01Sc, in which the vertex detector layers are 
approximated by cylinders.

Photon conversions were switched off in GEANT, as these can easily be 
suppressed later. Tracks stemming from hadronic interactions in the beam pipe 
and in the vertex detector layers were suppressed at the track selection stage 
using MC information. 

The LDCTracking package by A. Raspereza \cite{Raspereza} was used to simulate 
tracker hit digitization assuming simple Gaussian smearing of the hits, and for 
track reconstruction. Resulting tracks were fed into the Wolf particle flow 
algorithm \cite{Raspereza:2006kg} to obtain ReconstructedParticle objects, 
which were passed on to the Satoru jet finder \cite{Yamashita} to perform jet
finding using the Durham $kT$-cluster algorithm with a $y$-cut value of $0.04$. 

\begin{figure*}[h]
\centering
\mbox{
\includegraphics[width=0.80\columnwidth]{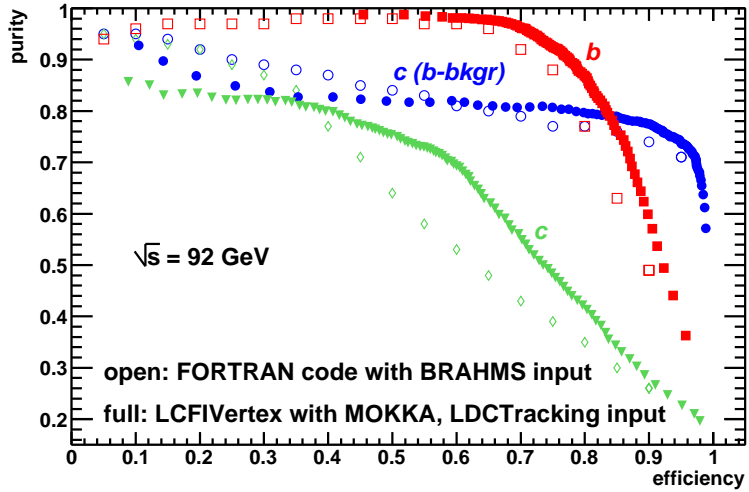}}
\caption{Comparison of tagging performance achieved with the LCFIVertex 
package and the previous FORTRAN code, using identical input events at the 
$Z$-resonance. For the FORTRAN case, events were passed through the BRAHMS 
simulation and reconstruction, for the LCFIVertex package the detector response 
was simulated using MOKKA and event reconstruction performed using MarlinReco. 
The new code, run with this input, yields better tagging performance, see text.}
\label{Fig:FullMCPurEff}
\end{figure*}

The default track selection used in the LCFIVertex package is based on a 
previous study \cite{Xella Hansen:2003sw} using the BRAHMS MC and reconstruction, 
with the modification that tracks stemming from the decays of K-shorts and Lambdas 
are suppressed using MC information.

Figure~\ref{Fig:FlavourTagInputs} shows the most sensitive input variables for the 
flavour tag as used in the Hawkings approach. Compared to the previous BRAHMS result 
\cite{Xella Hansen:2003sw}, these variables provide somewhat better separation 
power. The resulting flavour tagging performance, presented in 
Figure~\ref{Fig:FullMCPurEff}, is hence improved. Reasons for the difference seen may 
include a better detector resolution of $2\,\mu\mathrm{m}$ being assumed for the new 
result, compared to the former value of $3.5\,\mu\mathrm{m}$, as well as the 
suppression of K-short and Lambda-decays, photon conversions and hadronic 
interactions, all of which will need to be taken into account properly in a future 
version of the software.

The LDCTracking code can be run in different modes. The results shown in 
Figure~\ref{Fig:FlavourTagInputs} and Figure~\ref{Fig:FullMCPurEff} correspond to the 
``track cheater'', which uses MC information to assign tracker hits to the 
different tracks, run with only the Silicon-based detectors, i.e. vertex detector, 
Silicon intermediate tracker (SIT) and forward tracker (FTD). It was shown that when 
including the hits in the TPC and replacing the track cheater with an algorithm 
including realistic pattern recognition, the resulting tagging performance does not 
change significantly.

\begin{figure*}[h]
\centering
\mbox{
\includegraphics[width=0.80\columnwidth]{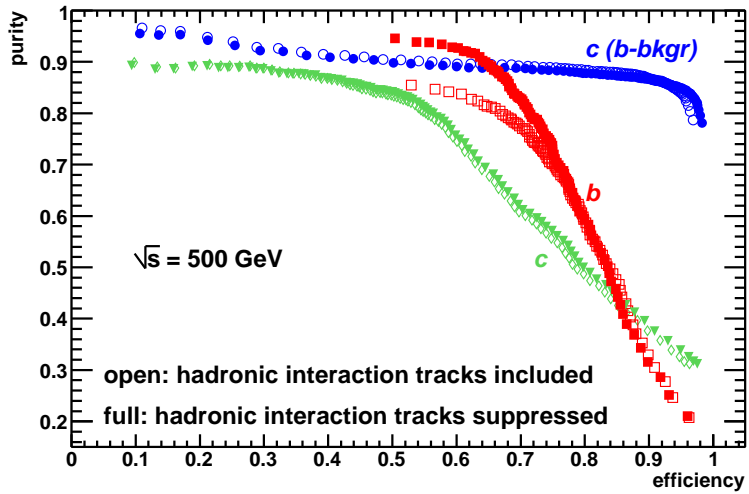}}
\caption{Effect of including tracks arising from hadronic interactions in the 
detector material and the beam pipe in the reconstruction, compared to the case, 
in which these tracks are suppressed using MC information. At a centre of mass 
energy of $\sqrt{s} = 500\,\mathrm{GeV}$, $b$-tagging performance clearly 
degrades.}
\label{Fig:HadronicInteractions}
\end{figure*}

The effect of including tracks arising from hadronic interactions in detector 
material and beam pipe is shown in Figure~\ref{Fig:HadronicInteractions}, compared 
to the current default settings of the package that suppress these tracks using MC 
information. At a centre of mass energy of $500\,\mathrm{GeV}$, the $b$-tag 
degrades significantly when tracks from hadronic interactions are included. At the 
$Z$-resonance this effect is negligible. Note that the current implementation of 
the suppression of the effect is based on MC information and only works for the 
specific detector model used for the initial full MC study (LDC01Sc). 

\section{Summary}

The LCFIVertex package provides the topological vertex finder ZVTOP, a flexible 
flavour tag with the Hawkings approach as default and determination of quark 
charge in heavy flavour jets. Validation of this C++ based code using the fast MC 
program SGV to simulate detector response and event reconstruction has shown it 
to be in good agreement with results from an earlier FORTRAN implementation, in 
comparison to which the new code has extended functionality, a higher degree of 
flexibility and improved documentation.

With input from the GEANT4-based detector simulation MOKKA and the event 
reconstruction package MarlinReco, the LCFIVertex package yields results comparable 
to those previously obtained from BRAHMS, with the differences being likely to be 
accounted for by a number of unrealistic simplifications made in the current first 
release version of the new code. One of these is the suppression of hadronic 
interaction effects using MC information, giving a clear improvement at high centre 
of mass energies.


\begin{footnotesize}

\end{footnotesize}



\begin{thebibliography}{99}
%
\bibitem{url} Slides: \\ 
\href{http://ilcagenda.linearcollider.org/contributionDisplay.py?contribId=76&sessionId=76&confId=1296}
{http://ilcagenda.linearcollider.org/contributionDisplay.py?contribId=76\&sessionId=76\&}
\href{http://ilcagenda.linearcollider.org/contributionDisplay.py?contribId=76&sessionId=76&confId=1296}
{confId=1296};
%
\bibitem{Jackson:1996sy}
  D.~J.~Jackson,
  Nucl.\ Instrum.\ Meth.\  A {\bf 388} (1997) 247;
%
\bibitem{Xella Hansen:2000dr}
  S.~M.~Xella Hansen, C.~Damerell, D.~J.~Jackson and R.~Hawkings,\\
  {\it Prepared for 5th International Linear Collider Workshop (LCWS 2000), Fermilab, Batavia, Illinois, 24-28 Oct 2000},
  \href{http://www-flc.desy.de/lcnotes/notes/LC-PHSM-2001-024.ps.gz}{LC-PHSM-2001-024};
%
\bibitem{Hillert:2005rp}
  S.~Hillert  [LCFI Collaboration],
{\it In the Proceedings of 2005 International Linear Collider Workshop (LCWS 2005), Stanford, California, 18-22 Mar 2005, pp 0313},
\href{http://www.slac.stanford.edu/econf/C050318/papers/0313.PDF}{ ECONF,C050318,0313};
%
\bibitem{Thom Wright}
  J.~Thom,
  ``Search for B/s0 - anti-B/s0 oscillations with a charge dipole  technique at
  SLD'', 
  DESY-THESIS-2002-006;\\
  T.~R.~Wright,
  ``Parity violation in decays of Z bosons into heavy quarks at SLD. ((C))'',
  UMI-30-33372;
%
\bibitem{Gaede:2003ip}
  F.~Gaede, T.~Behnke, N.~Graf and T.~Johnson,
{\it In the Proceedings of 2003 Conference for Computing in High-Energy and Nuclear Physics (CHEP 03), La Jolla, California, 24-28 Mar 2003, pp
TUKT001}
  \href{http://arxiv.org/pdf/physics/0306114}{arXiv:physics/0306114};
%
\bibitem{Wendt:2007iw}
  O.~Wendt, F.~Gaede and T.~Kramer,
  \href{http://arxiv.org/pdf/physics/0702171}{arXiv:physics/0702171};
%
\bibitem{org.lcsim}
  org.lcsim web page at \url{http://www.lcsim.org/software/lcsim/};
%
\bibitem{ILC software portal}
  ILC software portal at \url{http://ilcsoft.desy.de/portal};
%
\bibitem{Zeuthen CVS}
  Zeuthen CVS web interface \url{http://www-zeuthen.desy.de/lc-cgi-bin/cvsweb.cgi/};
%
\bibitem{Behnke:2001se}
  T.~Behnke and G.~A.~Blair,
  \href{http://www-flc.desy.de/lcnotes/notes/LC-TOOL-2001-005.ps.gz}{LC-TOOL-2001-005};
%
\bibitem{Behnke:2001qq}
  T.~Behnke, S.~Bertolucci, R.~D.~Heuer and R.~Settles,
  ECFA-2001-209;
%
\bibitem{SGV}
  SGV web page at \url{http://delphiwww.cern.ch/~berggren/sgv.html};
%
\bibitem{Adler:2006dw}
  V.~Adler,
  ``Optimising of design parameters of the TESLA vertex detector and search
  for events with isolated leptons and large missing transverse momentum  with
  the ZEUS-experiment (HERA II)'',
  DESY-THESIS-2006-012;
%
\bibitem{Jeffery}
  B.~Jeffery,
  {\it Contribution to International Linear Collider Workshop (ILC-ECFA and GDE Joint Meeting), Valencia, 6-10 November 2006,}  
  slides: 
\href{http://ilcagenda.linearcollider.org/getFile.py/access?contribId=72&sessionId=7&resId=5&materialId=slides&confId=1049}
{http://ilcagenda.linearcollider.org/getFile.py/}
\href{http://ilcagenda.linearcollider.org/getFile.py/access?contribId=72&sessionId=7&resId=5&materialId=slides&confId=1049}
{access?contribId=72\&sessionId=7\&resId=5\&materialId=slides\&confId=1049};
%
%
\bibitem{Sjostrand:2006za}
  T.~Sjostrand, S.~Mrenna and P.~Skands,
  JHEP {\bf 0605} (2006) 026
  \href{http://arxiv.org/pdf/hep-ph/0603175}{arXiv:hep-ph/0603175};
%
\bibitem{Mora de Freitas:2004sq}
  P.~Mora de Freitas,
{\it Prepared for International Conference on Linear Colliders (LCWS 04), Paris, France, 19-24 Apr 2004};
%
\bibitem{Raspereza}
  A.~Raspereza,
  ``LDC Tracking Software'',
  {\it in these proceedings};
%
\bibitem{Raspereza:2006kg}
  A.~Raspereza,
  \href{http://arxiv.org/pdf/physics/0601069}{arXiv:physics/0601069};
%
\bibitem{Yamashita}
  S.~Yamashita, 
  OPAL Technical Note TN579 (1998);
%
\bibitem{Xella Hansen:2003sw}
  S.~M.~Xella Hansen, M.~Wing, D.~J.~Jackson, N.~De Groot and C.~J.~S.~Damerell,
  ``Update on flavour tagging studies for the future linear collider using  the
  BRAHMS simulation'',
  \href{http://www-flc.desy.de/lcnotes/notes/LC-PHSM-2003-061.ps.gz}{LC-PHSM-2003-061}.
%
%
\end{thebibliography}
\end{document}